\documentclass[11pt.a4paper]{article}
\pdfoutput=1 % if your are submitting a pdflatex (i.e. if you have
\usepackage{jheppub} % for details on the use of the package, please
\usepackage[T1]{fontenc} % if needed

\title{
	Generalized Birkhoff theorem and its applications in mimetic gravity
}
\author[a]{Xin-zhou Li,}
\author[a,b]{Xiang-hua Zhai,}
\author[b]{Ping Li}
\affiliation[a]{Department of Mathematics, Shanghai Normal University, 100 Guilin Road, Shanghai 200234, China}
\affiliation[b]{Center for Astrophysics,  Shanghai Normal University, 100 Guilin Road, Shanghai 200234, China}

\emailAdd{kychz@shnu.edu.cn}
\emailAdd{zhaixh@shnu.edu.cn}
\emailAdd{lip57120@shnu.edu.cn}

\abstract{
There is undetermined potential function $V(\phi)$ in the action of mimetic gravity which should be resolved through physical means. In general relativity(GR), the static spherically symmetric(SSS) solution to the Einstein equation is a benchmark and its deformation also plays a crucial role in mimetic gravity. The equation of motion is provided with high nonlinearity, but we can reduce primal nonlinearity to a frequent Riccati form in the SSS case of mimetic gravity.  In other words, we obtain an expression of solution to the functional differential equation of motion with any potential function. Remarkably, we proved rigorously that there is a zero point of first order for the metric function $\beta(r)$ if another metric function $\alpha(r)$ possesses a pole of first order within mimetic gravity. The zero point theorem may be regarded as the generalization of Birkhoff theorem $\alpha\beta=1$ in GR. As a corollary, we show that there is a modified black hole solution for any given $V(\phi)$, which can pass the test of solar system. As another corollary, the zero point theorem provides a dynamical mechanism for the maximum size of galaxies. Especially, there are two analytic solutions which provide good fits to the rotation curves of galaxies without the demand for particle dark matter.
}
\begin{document}


\maketitle
\flushbottom

\section{Introduction}
The mimetic gravity is one of the particularly interesting theories of modified gravity which has emerged in the past few years. It is possible to describe the dark matter and dark energy of the universe as a purely geometrical effect, without the need of introducing additional dark components. The nature of dark matter is a real puzzle, which persistently evades any kind of detection outside the realm of gravitational interactions at galactic and cosmological scales. Chamseddine and Mukhanov \cite{Chamseddine1} introduced for the first time the concept of mimetic field. In Ref. \cite{Chamseddine1}, they showed a conformal extension of the general relativity (GR), in which the physical metric is defined in terms of an auxiliary metric and the first derivatives of a scalar field $\phi$ (mimetic field). An equivalent formulation of the mimetic dark matter theory was given in Ref. \cite{Golovnev1}, where the action employed a Lagrange multiplier as a constraint. Furthermore, this theory has been extended to a generalized version with the addition of an arbitrary potential \cite{Chamseddine2} and as a consequence, one can induce nearly any gravitational properties of the known substance including quintessence and phantom.  The ghost-free models and cosmological perturbation of mimetic gravity were discussed in Refs. \cite{Barvinsky} and \cite{Chamseddine3} , respectively. An interesting model has been proposed which does not only lead to mimetic dark matter but also provides a new approach to resolve singularities in GR \cite{Chamseddine4}. There exist several routes to mimetic gravity including the disformal transformation, Lagrange multiplier and singular Brans-Dicke theory \cite{Bekenstein,Deruelle}. With the use of these methods, various models have been proposed, for example, mimetic Horava gravity \cite{Cognola}, mimetic Horndeski gravity \cite{Achour} and mimetic $F(R)$ \cite{Odintsov}. Golovnev \cite{Golovnev2} has extended de Rham-Gabadadze-Tolley (dRGT) theory by a disformal transformation of the metric. Recently, Chamseddine and Mukhonov showed a ghost free mimetic massive gravity where the mass of graviton can be generated by using a Brout-Englert-Higgs mechanism with four scalar fields \cite{Chamseddine5}. Furthermore, they complete an explicit analysis using the methods of cosmological perturbation theory and consider quantum fluctuations of the massive graviton and mimetic field \cite{Chamseddine6}.

In GR, the static spherically symmetric (SSS) solution to the Einstein equation is a benchmark, and its massive deformation also plays a crucial role in massive gravity \cite{Li1,Lip}. Especially, we have found that there are seven black hole solutions in dRGT theory \cite{Li2}. In Refs. \cite{Myrzakulov1,Myrzakulov2}, the authors demonstrated how to reconstruct the potential for some interesting cases, including a correction to the Schwarzschild metric, a traversable wormhole, and so on. However, it is impossible to obtain an explicit expression for the potential $V(\phi)$ and exact metric solution using a so-called reconstructed method.

In this paper, we prove that there is a zero point for one SSS metric function and a pole of first order for another. There is always a modified black hole solution for any given $V(\phi)$ within mimetic gravity, which can pass the test of solar system. Using this zero point theorem, we provide a dynamical mechanism for the maximum size of galaxies. Furthermore, we find a universal functional expression of SSS solution for any potential $V(\phi)$ in mimetic gravity, and we reduce primal high nonlinearity of the Einstein equation to a frequent Riccati's form of first differential equation. Using this formula, we find some solutions that can be considered as candidates of black hole, or can explain the flat rotation curves of spiral galaxies within the mimetic gravity. The zero point theorem occupied a key position in this paper, which can be regarded as a generalization of Birkhoff theorem in GR. Especially, this theorem predicts that the product of two metric functions is not 1 as in GR but it is a regular function.

\section{Functional expression of SSS solution}
\subsection{The equations of motion under SSS ansatz}

In a cosmological context, the mimetic field plays the role of a "clock". Thus one can fancy making the mimetic field dynamical by adding a potential $V(\phi)$. The action of mimetic gravity is
\begin{equation}\label{action}
S=\int d^4 x\sqrt{-g}[R+\lambda(1-g^{\mu\nu}\partial_\mu \phi \partial_\nu \phi)+V(\phi)+\mathcal{L}_m],
\end{equation}
where $R$ is the Ricci scalar, $\lambda$ is a Lagrange multiplier, $\mathcal{L}_m$ is the Lagrangian of usual matter and we set $16\pi G=1$. Variation of action (\ref{action}) with respect to the metric gives the following equations
\begin{equation}\label{einsteintensor}
G_{\mu\nu}=R_{\mu\nu}-\frac 1 2g_{\mu\nu}=\tilde{T}_{\mu\nu}+T_{\mu\nu},
\end{equation}
where $T_{\mu\nu}$ is the energy-momentum tensor of the usual matter and
\begin{equation}
\tilde{T}_{\mu\nu}=\lambda\partial_\mu \phi\partial_\nu\phi+\frac {\lambda}2(1-g^{\alpha\beta}\partial_\alpha\phi\partial_\beta\phi)g_{\mu\nu}+\frac V 2 g_{\mu\nu},
\end{equation}
which describes the extra contribution to Einstein equations due to the $\phi$-dependent terms. Variation with respect to $\phi$ gives the motion equation of mimetic scalar field as
\begin{equation}\label{scalarequation}
\partial_\nu(\sqrt{-g}\lambda g^{\mu\nu}\partial_\mu \phi)+\frac 1 2 \sqrt{-g}V_\phi=0.
\end{equation}
Obviously, the mimetic scalar field $\phi$ satisfies the constraint
\begin{equation}\label{constraint}
g^{\mu\nu}\partial_\mu\phi\partial_\nu\phi=1.
\end{equation}
Taking the trace of (\ref{einsteintensor}), we obtain
\begin{equation}
\lambda=G-2V-T,
\end{equation}
and (\ref{einsteintensor}) can be rewritten as
\begin{equation}\label{einsteinequation}
G_{\mu\nu}-(G-2V-T)\partial_\mu\phi\partial_\nu\phi-\frac V 2 g_{\mu\nu}=T_{\mu\nu}.
\end{equation}
By using the constraint (\ref{constraint}), $\nabla^\mu G_{\mu\nu}=0$ and $\nabla^\mu T_{\mu\nu}=0$, we can rederive the mimetic scalar equation (\ref{scalarequation}).

Next, we consider the SSS ansatz as follows
\begin{equation}\label{ansatz}
ds^2=-\beta(r)dt^2+\alpha(r)dr^2+r^2d\Omega^2,
\end{equation}
and $\phi=\phi(r)$. In the SSS case, the constraint (\ref{constraint}) can be reduced to
\begin{equation}\label{reducedconstraint}
\phi'^2=\alpha.
\end{equation}
Thus, we obtain the equations of motion under aforesaid ansatz,
\begin{equation}\label{system1}
\frac 1 {r^2}\left(1-\left(\frac r \alpha\right)'\right)+\frac 1 2 V=0,
\end{equation}
\begin{equation}\label{system2}
\frac 1 {r^2}\left(1-\frac {(\beta r)'} {\alpha\beta}\right)-\lambda+\frac 1 2 V=0,
\end{equation}
\begin{equation}\label{system3}
(\sqrt{\beta}r^2\lambda)'+\frac 1 2\sqrt{\alpha\beta}r^2V_\phi=0,
\end{equation}
where the primes denote derivatives with respect to $r$. We would show that presetting the potential function $V(r)$ is equivalent to setup of $V(\phi)$. In reality, we can find $\alpha$ from (\ref{system1}) and then obtain $\phi(r)$ by the integral of $\sqrt{\alpha(r)}$ for any potential function $V(r)$ and $T^0\hspace{0.1cm}_0(r)$. Combining $V(r)$ with $\phi(r)$, we obtain the potential $V(\phi)$ easily. Furthermore, we have $V_\phi=\sqrt{\alpha}V'$ from (\ref{reducedconstraint}). Thus, (\ref{system1})-(\ref{system3}) are the system containing three differential equations if and only if we take a fixed function $V(r)$. From this point of view, the system contains three equations for three unknowns, $\alpha, \beta$ and $\lambda$. It is solvable even though it is a highly nonlinear coupling system.

\subsection{Reduction of nonlinearity theorem}
Meanwhile, we can regard (\ref{system1})-(\ref{system3}) as functional differential equations on the potential $V(\phi)$. We adopt a strategy as follows: The first step is to find the relation between $\alpha(r), \phi(r)$ and $V(\phi)$. Let $\alpha^{-1}=1-\frac{r_s}r+\Delta(r)$ and using (\ref{reducedconstraint}), we have
\begin{equation}\label{Delta}
\Delta(r)=\frac 1{2r}\int^r V\rho^2d\rho,
\end{equation}
and
\begin{equation}\label{potential0}
V(r)=\frac{2\Delta'(r)}r+\frac{2\Delta(r)}{r^2}.
\end{equation}
From (\ref{reducedconstraint}), we obtain
\begin{equation}
\phi(r)=\int^r \frac{d\rho}{\sqrt{1-\frac{r_s}\rho+\Delta(\rho)}}\equiv\mathbb{I}(r),
\end{equation}
where $\mathbb{I}$ is an integral operator and $r=\mathbb{I}^{-1}(\phi)$. Thus,
\begin{equation}
V(\phi)=\frac{2\Delta'(\mathbb{I}^{-1}(\phi))}{\mathbb{I}^{-1}(\phi)}+\frac{2\Delta(\mathbb{I}^{-1}(\phi))}{\left[\mathbb{I}^{-1}(\phi)\right]^2}.
\end{equation}
Second, using (\ref{system1}) and (\ref{system2}) we have
\begin{equation}\label{lambda}
\lambda(r)=\mp\frac 1 {\sqrt{\beta}r^2}\int^r\frac{V'}{\sqrt{\alpha}}d\rho.
\end{equation}
Third, we derive $\beta(r)$ equation from (\ref{system3}) and (\ref{lambda}) as follows
\begin{equation}\label{betaequation}
\left(\sqrt{\beta}\right)''+\left(\frac 1 r-\frac 1 2\frac{\alpha'}\alpha\right)\left(\sqrt{\beta}\right)'+\left[\frac 1 {r^2}(\alpha-1)+\frac{\alpha'}{2r\alpha}\right]\sqrt{\beta}=0.
\end{equation}

Though (\ref{reducedconstraint})-(\ref{system3}) compose a system with high nonlinearity at first glance, we can reduce it by certain arithmetic. In the case of constant potential, (\ref{betaequation}) will becomes a linear equation of first order
\begin{equation}
\left(\sqrt{\beta}\right)'+\frac 1 2\left(\mathrm{ln} \alpha\right)'\sqrt{\beta}=\frac{\mu\alpha}r,
\end{equation}
and its exact solution is
\begin{equation}\label{solutionB}
\beta=\frac 1 \alpha\left(\mu\int^r\frac{\alpha^{\frac 3 2}d\rho}\rho+\nu\right)^2,
\end{equation}
where $\mu$ and $\nu$ are integral constants. In the case of nonconstant potential, let $v=\beta'/2\beta$, we can transfer (\ref{betaequation}) into a Riccati equation
\begin{equation}\label{Riccati}
v'+v^2+(\frac 1 r-\frac 1 2 \frac{\alpha'}{\alpha})v=\frac 1 {r^2}(1-\alpha)-\frac 1 {2r}\frac{\alpha'}{\alpha},
\end{equation}
and the metric function
\begin{equation}\
\beta=\frac{|\alpha|}{\alpha}\left(\mu\mathrm{exp}\left[\int^r v(\rho)d\rho\right]\right)^2, \label{beta}
\end{equation}
where $\mu$ is an integral constant.

There is not a general method for solving Riccati equation. Nevertheless, when a special solution $v_0(r)$ is known by guess or observation, then one can write the general solution in the form
\begin{equation}
v=v_0+\frac 1 u,
\end{equation}
where
\begin{equation}
u(r)=\exp\left[\int^r p(\rho)\rho\right]\left(\int^r \exp\left[-\int^{\xi}p(\rho)d\rho\right]d\xi\right),
\end{equation}
and
\begin{equation}
p(r)=\frac 1 r+2v_0-\frac 1 2 \frac{\alpha'}{\alpha}.
\end{equation}
In summary, (\ref{Delta}), (\ref{lambda}), (\ref{solutionB}) and (\ref{beta}) make up the functional expressions of SSS solution, and the primal highly nonlinear problem has been reduced to Riccati's form. Thus, we have the following theorem

\textbf{Theorem 1 (Reduction of nonlinearity theorem).}
\emph{In the case of constant potential, the high nonlinearity of equations of motion will fade away so that it is exact solvable; In the case of nonconstant potential, this nonlinearity can be reduced to Riccati's form so that the problem of exact solution becomes the resolution of a certain Riccati equation.
}

\subsection{A pedagogical example}

Using theorem 1, we can find a series of SSS solutions for various potential. We take the mimetic bouncing potential as a pedagogical instance of using the formulae above. The bouncing potential is
\begin{equation}\label{potential}
V(\phi)=\frac{2s^2}{\left[(\phi-\phi_0)^2+s^2\right]}.
\end{equation}
Taking $\Delta(r)=-\frac{s^2}{r^2}$, we have $V(r)=\frac{2s^2}{r^4}$ from (\ref{potential0}). In the two asymptotic cases, we can obtain the explicit expression of $V(\phi)$. For the case of $r\ll \frac{s^2}{r_s}$, we have $\mathbb{I}(r)=\sqrt{r^2-s^2}+\phi_0$ and bouncing potential (\ref{potential}) is sure to recur. For the case of $r\gg \frac{s^2}{r_s}$, \emph{i.e.}, $\Delta(r)\approx0$, then $V(\phi)\approx0$. Therefore, we have
\begin{equation}
V(\phi)=\frac{2s^2}{\left[\mathbb{I}^{-1}(\phi)\right]^4}=\begin{cases}\frac{2s^2}{\left[(\phi-\phi_0)^2+s^2\right]^2},\\\\0,\end{cases} \text{for}
\begin{array}{c}r\ll \frac{s^2}{r_s},\\\\r\gg \frac{s^2}{r_s}.\end{array}
\end{equation}
Since $\Delta(r)=-\frac{s^2}{r^2}$, $\alpha(r)$ contains a pole of first order
\begin{equation}
r_h=\frac 1 2 (r_s+\sqrt{r_s^2+4s^2}).
\end{equation}
Thus, we obtain the solution of (\ref{betaequation}) as follows
\begin{equation}\label{metricfunction}
\beta(r)=\begin{cases}-\mu^2(\frac 2 \pi \sin \frac r s-1)^2,\\\\\mu^2(\frac 2 \pi \mathrm{arccosh}\frac r s)^2,\\\\1-\frac{r_s}r,\end{cases} \text{for}
\begin{array}{c}r<r_h,\\\\r_h<r\ll\frac{s^2}{r_s},\\\\r\gg\frac{s^2}{r_s},\end{array}
\end{equation}
From (\ref{lambda}), we have
\begin{equation}
\lambda(r)=\frac{\mu}{\sqrt{\beta}r^2}\begin{cases}\int^r \frac{8s^2(\rho^2-s^2)^{\frac 1 2}}{\rho^6}d\rho,\\\\0,\end{cases}  \text{for} \begin{array}{c}r\ll\frac{S^2}{r_s},\\\\r\gg\frac{S^2}{r_s}.\end{array}
\end{equation}
Obviously, this solution will pass the test of solar system if we take suitable parameter $s$.

The metric functions (\ref{metricfunction}) give surely the description of a black hole. It is known to all that the scalar $I=R_{\alpha\beta\gamma\delta}R^{\alpha\beta\gamma\delta}=12r_s^2/r^6$, so $r_s$ is only a coordinate singularity and $r=0$ is really a geometric singularity for the Schwarzschild metric. In the case of metric (\ref{metricfunction}), we have
\begin{equation}
I=\frac{12s^4}{r^6}+\begin{cases}\frac{12(s^2-r^2)}{r^6\left(\frac \pi 2 -\mathrm{arcsin} (\frac r s)\right)^2},\\\\\frac{12(r^2-s^2)}{r^6\left(\mathrm{arccosh} (\frac r s)\right)^2},\end{cases} \text{for} \begin{array}{c}r<r_h,\\\\r_h<r\ll\frac{s^2}{r_s},\end{array}
\end{equation}
which is similar to Schwarzschild one. Because all worldlines within the future light cone of objects terminate on $r=0$, the crash with the singularity cannot be averted.

Using the same procedure as in subsection 2.3 via choosing different potential $V(\phi)$, we can find new analytical solutions that are divided into two types according to the number of zero points for $\beta(r)$. Type I solutions have one zero point and the others are called Type II. The solution (\ref{metricfunction}) is an outstanding example of Type I. Type I may be considered as candidates of black holes in mimetic gravity. Type II will provide a good fit to the rotation curves of spiral galaxies within the mimetic gravity.

\section{Some universal properties of SSS solution}
\subsection{The zero point theorem}

Next, we show a universal property of SSS solution within mimetic gravity which may be regard as generalization of Birkhoff theorem in GR. We will prove that $\alpha\beta$ is an analytic function in mimetic gravity whereas $\alpha\beta=1$ in GR. $\alpha(r)$ is bound to a pole of first order since $\alpha^{-1}=1-\frac{r_s}r+\Delta(r)$, so we have
\begin{equation}
\alpha(r)=h(r)(r-r_h)^{-1},
\end{equation}
where $h(r)$ is a positive-definite and analytic function. According to the general theory of differential equation, $r=r_h$ is a regular singularity of (\ref{betaequation}) so that the local solution of (\ref{betaequation}) can be written as
\begin{equation}
\beta(r)=(r-r_h)^{2\rho}\left[\sum_{n=0}^{\infty}C_n(r-r_h)^n\right]^2,
\end{equation}
where $C_0\neq 0$ and $\rho$ is determined by its index equation. Let $\alpha_{-1}$ be the residual of $\alpha(r)$, we have
\begin{equation}
\lim_{r\rightarrow r_h}(r-r_h)\alpha = \alpha_{-1},\lim_{r\rightarrow r_h}\left(\frac 1 \alpha\right)' = \frac 1 {\alpha_{-1}}.
\end{equation}
Thus, the index equation is $\rho(\rho-1)+\frac 1 2\rho=0$ and $\rho=\frac 1 2$, and $C_n$ satisfy the recursion formula as follows
\begin{equation}
C_n=-\frac{\sum_{k=1}^n[(n+\frac 1 2-k)a_k+b_k]C_{n-k}}{n(n+\frac 1 2)},
\end{equation}
where $a_k$ and $b_k$ are defined by
\begin{eqnarray}
\sum a_k(r-r_h)^k &=& \frac{(r-r_h)}r\left[1+\frac 1 2 \alpha r(\frac 1 \alpha)'\right],\nonumber\\
\sum b_k(r-r_h)^k &=& \frac{(r-r_h)^2}{r^2}\left[\alpha-1-\frac r 2 \alpha(\frac 1 \alpha)'\right].
\end{eqnarray}
Clearly, $r=r_h$ is a zero point of first order of $\beta(r)$. We now have proven theorem 2 that is stated as follows:

\textbf{Theorem 2 (Zero point theorem).}
\emph{In mimetic gravity, there is a metric  function solution $\alpha(r)$ with a pole of first order at $r=r_h$ for a corresponding potential $V(\phi)$, but $\beta(r)$ will possess a zero point of first order at $r=r_h$. Therefore, $\alpha\beta$ will be regular.}

\subsection{Existence theorem of black holes}
As applications of the zero point theorem, we will obtain some physical corollaries in the small (star) scale and intermediate (galaxy) scale. In the small scale, the sign of $\beta(r)$ will change from positive to negative as $r$ decreases and crosses $r_h$ since $\beta(r)$ has a zero point at $r=r_h$. Let us consider a light signal propagating in the radial direction. The velocity of this signal is
\begin{equation}
\frac{dr}{dt}=\pm (r-r_h)\left[\frac{\sum_{n=0}^\infty C_n(r-r_h)^n}{h(r)}\right]^{\frac 1 2}.
\end{equation}
In the exterior, the axis of the light cones is parallel to the $t$-axis. In the interior, the axis is parallel to the $r$-axis and $r$ becomes a timelike coordinate. When $r$ approaches $r_h$ from outside, the lightcones become very narrow. When $r$ just crosses $r_h$ to the interior, the lightcones suddenly become very broad again, and their timelike regions come to be horizontal, so that the only possible directions of radial motion are towards the singularity $r=0$. Thus, we have the following theorem:

\textbf{Theorem 3.} \emph{On every account, there are solutions with event horizon for any given potential function $V(\phi)$ in mimetic gravity, which can be considered as candidates of black holes.}

\section{The zero point theorem applied to galaxies}
\subsection{A dynamical mechanism for the maximum size of galaxies}
If $\alpha$ has two poles of first order, $r_{in}$ and $r_{out}$, there are zero points of first order for $\beta(r)$ from the zero point theorem. That is to say, $\beta(r_{in})=\beta(r_{out})=0$ and $\beta(r)>0, r\in(r_{in}, r_{out})$ so that there exists a maximum $r_{max}$ between $r_{in}$ and $r_{out}$. And $r_{max}$ is the turning point of the Newtonian potential $\Phi(r)=\frac 1 2 (\beta-1)$. In other words, $\Phi(r)$ changes from attractive to repulsive at $r_{max}$ which provides a dynamical mechanism for the maximum size of galaxies. Thus, $r_{max}$ is just the maximum size of galaxies since the Newtonian potential $\Phi$ becomes repulsive at $r>r_{max}$. As a corollary of the zero point theorem, we have the following theorem:

\textbf{Theorem 4.} \emph{The Newtonian potential will change from attractive to repulsive between two poles if $\alpha(r)$ has two poles of first order. This mathematical result provides a dynamical mechanism for the maximum size of galaxies.}

This theorem requires that $\beta(r)$ has two zero points at least, so such solutions should belong to Type II.

On the other hand, it is common knowledge that the rotational velocity of spiral galaxies does not fall off as expected in GR with only the luminous matter as source. Mannheim \textit{et. al.} \cite{Mannheim1,Mannheim2} exhibited a model which provides an explanation for the inferred flat rotation curves of spiral galaxies within the conformal gravity framework \cite{Mannheim3,Zhang}. The authors of Ref.\cite{Cognola} showed the similar pattern in mimetic gravity framework. In this paper, we shall find two solutions which could apply to explain the inferred flat rotation curves of spiral galaxies.

\subsection{Type I solution for the rotation curves of spiral galaxies}
We consider the potential $V(\phi)$ with an intermediate scale parameter $\gamma_0$ and its asymptotic form is
\begin{equation}
V(\phi)=\begin{cases}0,\\\\\frac{16}{(\phi-\phi_0)^2-4\gamma_0^2},\end{cases}\quad \text{for} \quad \begin{array}{c}r\ll \gamma_0,\\\\r\gg r_s,\end{array}
\end{equation}
where $r_s$ is the event horizon of solution. Thus, we have $\phi=\gamma_0(1+r/\gamma_0)^{\frac 1 2}+\phi_0$ and $V(r)=4/\gamma_0 r$. Furthermore, the metric functions are
\begin{equation}
\alpha=\begin{cases}(1-\frac{r_s}r)^{-1},\\\\(1+\frac r{\gamma_0})^{-1},\end{cases}\quad \text{for} \quad \begin{array}{c} r \ll\gamma_0,\\\\ r \gg r_s,\end{array}
\end{equation}
\begin{equation}
\beta=\begin{cases}1-\frac{r_s}r,\\\\(1+\frac 3 2 \frac r{\gamma_0})^2,\end{cases}\quad \text{for} \quad \begin{array}{c} r \ll\gamma_0,\\\\r \gg r_s.\end{array}
\end{equation}
On sufficiently large scales, the Newtonian potential $\Phi\simeq\frac{c^2 r}{2\gamma_0}$ where the speed of light $c$ has been resumed. It is easy to see that the rotation velocity profile increases slightly as $\sqrt{r}$. This solution is fascinatingly in accordance with astrophysical data for small and medium sized low surface brightness (LSB) galaxies. However, the high surface brightness (HSB) galaxies are quite different from LSB, where the Newtonian contribution might be sufficient to complete with the rising linear term. When the rotation velocity departs enough far from the center of HSB, its rising behavior is arrested. For this, We have to turn our attention to Type II solution.

\subsection{Type II solution for the rotation curves of spiral galaxies}
We consider the potential $V(\phi)$ with two intermediate scale parameters $\gamma_0$ and $\lambda_0$, and its asymptotic form is
\begin{equation}
V(\phi)=\begin{cases}0,\\\\-\frac 6{\gamma_0}+\frac 8{\sqrt{\mu}\sin\sqrt{\frac 1{\lambda_0}}(\phi-\phi_0)+\frac 1 {\gamma_0}},\end{cases} \text{for}\begin{array}{c}r\ll \gamma_0,\lambda_0,\\\\r\gg r_s,\end{array}
\end{equation}
where $\mu=4\gamma_0^2\lambda_0+\lambda_0^2$, and $r_s$ is the event horizon of the solution. Thus, we have
\begin{equation}
\phi(r)=\sqrt{\lambda_0}\arcsin\left(\frac{2\gamma_0 r-\lambda_0}{\sqrt{\mu}}\right)
\end{equation}
and $V(r)=-\frac 6 {\gamma_0}+\frac 4{\lambda_0 r}$. Furthermore, the metric functions are
\begin{equation}\label{alpha}
\alpha=\begin{cases}(1-\frac{r_s}r)^{-1},\\\\(1+\frac r{\gamma_0}-\frac{r^2}{\lambda_0})^{-1},\end{cases}\quad \text{for} \quad \begin{array}{c} r \ll\gamma_0,\lambda_0\\\\ r \gg r_s,\end{array}
\end{equation}
\begin{equation}
\beta=\begin{cases}1-\frac{r_s}r,\\\\\sum_{k=0}^\infty(\sum_{i+j=k}a_ia_j)r^k,\end{cases}\quad \text{for} \quad \begin{array}{c} r \ll\gamma_0,\lambda_0\\\\r \gg r_s.\end{array}
\end{equation}
where $a_0=1, a_1=\frac 3 {2\gamma_0}, a_2=-\frac 1 {2\lambda_0}$ and the recursion formula is
\begin{equation}\label{recursion}
a_{k+1}=\frac{(k-2)(k+1)\lambda_0a_{k-1}-(k-1)(k+\frac 3 2)\gamma_0a_k}{\gamma_0\lambda_0(k+1)^2}.
\end{equation}
From (\ref{alpha}-\ref{recursion}) and $r\gg r_s$, we have
\begin{eqnarray}
r_{out}&=&\frac 1 2\left(\frac{\lambda_0}{\gamma_0}+\sqrt{\left(\frac{\lambda_0}{\gamma_0}\right)^2+4\lambda_0}\right),\nonumber\\
r_{max}&=&\frac 3 2\left(\frac {\gamma_0}{\lambda_0}-\frac 9 {\gamma_0}\right)^{-1}.
\end{eqnarray}
At the first blush, we should fit the values of two parameters $\gamma_0$ and $\lambda_0$ using the data from rotation curves. In reality, it is possible that one can adopt the results of Refs.\cite{Mannheim2,OBrien} where the same pattern were fitted to rotation curves as the higher order terms are neglected by using the recursion relation (\ref{recursion}). The total sample is composed of 138 galaxies where 25 galaxies are dwarf galaxies and 21 galaxies have data points that are enough far from the optical disk region. We obtain the following results with the aid of Ref.\cite{Mannheim2},
\begin{equation}
\gamma_0=9.80\times 10^{19}\textmd{cm}, \quad \lambda_0=1.05\times 10^{53}\textmd{cm}^2,
\end{equation}
therefore, it is reasonable that the higher order terms are neglected. Furthermore, we have
\begin{equation}
r_{out}=3.23\times 10^{26}\textmd{cm}, \quad r_{max}=4.81\times 10^{23} \textmd{cm}.
\end{equation}
Thus, this Type II solution shows that our dynamical mechanism(Theorem 4) is indeed effective for the inferred flat rotation curves of spiral galaxies within the mimetic gravity framework, without the need for particle dark matter.

\section{Summary}
It is a universal property that there is a zero point of first order for the SSS metric function $\beta(r)$ if another metric function $\alpha(r)$ possesses a pole of first order within mimetic gravity. The zero point theorem occupied a key position in this work, which can be regarded as a generalization of Birkhoff theorem in GR. As its corollary, we show that there is a modified black hole solution for any given $V(\phi)$, which can pass the test of solar system. Furthermore, the zero point theorem provides a dynamical mechanism for the maximum size of galaxies. There are two analytic solutions which give good fits to the rotation curves of spiral galaxies without the demand for particle dark matter.

Finally, we give a brief discussion for the cosmological horizon. The visible universe is a spherical region centered on us, from within which signal of gravitational wave has had time to reach us since the universe began. The boundary of our visible universe is called our horizon. An ideal result should satisfy that $r_{out}$ is near our horizon. Maybe we need some improved solutions. For example, $\alpha(r)=1+\frac r {\gamma_0}-\frac{r^{2-\epsilon}}{\lambda_0}$, for $\gamma_0,\lambda_0\gg r_s$ and $\epsilon$ is a small and positive constant. We will study it in our future work.

\end{document}